\overfullrule=0pt
\input harvmac
\def\a{{\alpha}}
\def\ah{{\hat\alpha}}
\def\bh{{\hat\beta}}

\def\l{{\lambda}}
\def\lh{{\hat\lambda}}
\def\th{{\hat\theta}}
\def\b{{\beta}}
\def\g{{\gamma}}
\def\d{{\delta}}

\def\e{{\epsilon}}
\def\s{{\sigma}}

\def\L{{\Lambda}}
\def\O{{\Omega}}

\def\half{{1\over 2}}
\def\p{{\partial}}

\def\pt{{{\partial\over{\partial\tau}}}}
\def\t{{\theta}}

\Title{\vbox{\hbox{ }}}
{\vbox{
\centerline{\bf Pure Spinors, Twistors, and Emergent Supersymmetry}}}
\bigskip\centerline{Nathan Berkovits\foot{e-mail: nberkovi@ift.unesp.br}}
\bigskip
\centerline{\it Instituto de F\'\i sica Te\'orica, Universidade Estadual
Paulista}
\centerline{\it R. Bento T. Ferraz 271, 01140-070, S\~ao Paulo, SP, Brasil}
\centerline{and}
\centerline{\it Kavli Institute for Theoretical Physics, University of
California}
\centerline{\it Santa Barbara, CA 93106, USA}

\vskip .3in
Starting with a classical action whose matter variables are a d=10
spacetime
vector $x^m$ and a pure spinor $\l^\a$, 
the pure spinor formalism for the superstring
is obtained by gauge-fixing the twistor-like constraint
$\partial x^m (\gamma_m \lambda)_\alpha =0$.
The fermionic variables
$\theta^\alpha$ are Faddeev-Popov ghosts coming from this gauge-fixing and
replace the usual $(b,c)$ ghosts coming from gauge-fixing the
Virasoro constraint.
After twisting the ghost-number such that 
$\theta^\alpha$ has ghost-number zero and $\lambda^\alpha$ has ghost-number
one, the BRST cohomology describes the usual 
spacetime supersymmetric states of the superstring.

\Date {May 2011}

\newsec{Introduction}

For computing multiloop scattering amplitudes or constructing quantizable
sigma models in Ramond-Ramond backgrounds, the pure spinor formalism is
the most convenient description of the superstring \ref\pureor
{N. Berkovits, {\it Super-Poincare covariant quantization of the
superstring}, JHEP 0004 (2000) 018, hep-th/0001035.}. This formalism contains
the usual $(x^m,\t^\a)$ matter variables of d=10 superspace, as well as bosonic
ghost variables $\l^\a$ satisfying the pure spinor constraint $\l\g^m\l=0$.
Physical states are elegantly described by the ghost-number one
cohomology of the BRST operator $Q=\int \l^\a d_\a$ where $d_\a$ is the
fermionic Green-Schwarz constraint. 

Since $\{d_\a,d_\b\}=\g^m_{\a\b} \Pi_m$ where $\Pi_m$ is the supersymmetric
momentum, $d_\a$ contains both first and second-class constraints. So the 
nilpotence of $Q$ does not follow from the Jacobi identity of a first-class
constraint algebra but instead requires that the ghost variable $\l^\a$ satisfies
the pure spinor constraint. This unusual feature has made it difficult to obtain
$Q$ from gauge-fixing a classical action using the standard
BRST procedure.
Another unusual feature of $Q$ in the pure spinor formalism is that it does not
involve $(b,c)$ worldsheet reparameterization
ghosts or the Virasoro constraint.

In this paper, a new interpretation of $Q$ will be proposed which 
elegantly explains its origin. Instead of intepreting $(x^m,\t^\a)$ as
matter variables and $\l^\a$ as ghost variables, 
$x^m$ and $\l^\a$ will be interpreted as matter variables 
and $\t^\a$ as ghost variables. And instead of imposing the usual Virasoro
constraint $\p x^m \p x_m + ... =0$, the matter
variables will instead satisfy $\p x^m (\g_m \l)_\a=0$. 
This constraint implies that $\p x^m = \l\g^m h$ for some spinor $h^\a$,
so it is a ten-dimensional version of the d=4 twistor constraint 
$\p x^m = \l^a \s^m_{a\dot a} \bar\l^{\dot a}$.\ref\puretw{
L.P. Hughston, {\it The Wave Equation in Even Dimensions}, Further Advances in
Twistor Theory, vol. 1, Research Notes in Mathematics 231, Longman, pp. 26-27, 1990\semi
 L.P. Hughston, {\it A Remarkable Connection between the Wave Equation and Pure Spinors in Higher Dimensions}, Further Advances in Twistor Theory, vol. 1, Re- search Notes in Mathematics 231, Longman, pp. 37-39, 1990\semi
L.P. Hughston and L.J. Mason, {\it A Generalized Kerr-Robinson Theorem}, 
Classical
and Quantum Gravity 5 (1988) 275.}
\ref\cherkis{N. Berkovits and S. Cherkis, {\it Higher-dimensional
twistor transforms using pure spinors}, JHEP 0412 (2004) 049,
hep-th/0409243\semi N. Berkovits, {\it Ten-dimensional super-twistors
and super-Yang-Mills}, JHEP 1004 (2010) 067, arXiv:0910.1684.}
Quantizing this twistor-like constraint in the standard manner
leads to fermionic Faddeev-Popov
ghosts $\t^\a$ and a nilpotent BRST operator. After twisting the ghost-number so
that $\t^\a$ has ghost-number zero and $\l^\a$ has ghost-number one,
the cohomology of this BRST operator will be related to
the cohomology of $Q=\int \l^\a d_\a$.

Although the classical action has no fermionic variables, the BRST
cohomology is spacetime supersymmetric after twisting the ghost charge.
The ``emergence'' of d=10 spacetime supersymmetry after imposing a twistor-like
constraint is quite surprising and may explain why twistor methods 
have been so useful for describing spacetime supersymmetric theories.
Although twistors were originally developed by Penrose to describe 
purely bosonic d=4 theories, their most powerful applications have been for
d=4 theories with maximal spacetime supersymmetry.

Another interesting observation
is that a projective pure spinor $\l^\a$ parametrizes
$SO(10)/U(5)$ which chooses a complex structure of
the d=10 (Wick-rotated) spacetime. So 
the twistor-like constraint $\p x^m (\g_m\l)_\a=0$ resembles the constraint
$\p x^I=0$
of a $\hat c=5$ topological string where $I=1$ to 5 labels
the holomorphic directions. As pointed out by Nekrasov\ref\nekrasov{N. Nekrasov,
KITP lecture ``Pure spinors, beta-gammas, super-Yang-Mills and
Chern-Simons'',
http://online.kitp.ucsb.edu/online/strings09/nekrasov2/ , January 2009.}, 
introducing $\l^\a$ to dynamically determine
the complex structure enlarges the cohomology of the topological string BRST
operator to the full superstring spectrum.

\newsec{Superparticle}

\subsec{Worldline action and BRST operator}

Before discussing this gauge-fixing procedure for the superstring, it will
be useful to first discuss the procedure for the superparticle whose spectrum
is d=10 super-Yang-Mills. For the superparticle, the classical worldline action
will be defined as 
\eqn\superp{S = \int d\tau [ P_m \pt
 x^m + w_\a \pt \l^\a + f^\a (P_m \g^m\l)_\a]}
where $m=0$ to 9, $\a=1$ to 16, $\l^\a$ is a pure spinor
satisfying 
\eqn\pures{\l\g^m\l=0,} and
$f^\a$ is a Lagrange multiplier for the twistor-like
constraint $P_m (\g^m\l)_\a=0$. 

The first step is to gauge-fix the Lagrange multiplier $f^\a =0$ which
introduces the fermionic Faddeev-Popov ghosts $(\t^\a, p_\a)$ with worldsheet
action $\int d\tau p_\a \pt \t^\a$.
The resulting BRST operator is $Q = \t^\a  P_m (\g^m \l)_\a$ where $(\t^\a,p_\a)$
carry ghost-number $(+1,-1)$.

Since only 5 of the 16 components of the constraint $P_m (\g^m\l)_\a$
are independent, there are bosonic
ghosts-for-ghosts coming from the gauge transformation
of the Lagrange multiplier
$\d f^\a = \e^\a$ where $\e^\a$ satisfies $\e^\a\g^m_{\a\b}\l^\b=0$.
This implies the introduction of bosonic ghost-for-ghosts
$(u^\a, v_\a)$ with worldsheet action $\int d\tau v_\a\pt u^\a$
where $u^\a$ is constrained to satisfy 
\eqn\ucon{u^\a \g^m_{\a\b} \l^\b =0}
and
$(u^\a, v_\a)$ 
carry ghost-number $(+2,-2)$.
The resulting BRST operator
is 
\eqn\brstsuperp{Q = \t^a P_m (\g^m \l)_\a + u^\a p_\a}
and the gauge-fixed worldline action is 
\eqn\superpf{S = \int d\tau [ P_m \pt
x^m + w_\a \pt \l^\a + 
p_\a \pt \t^\a + 
v_\a\pt u^\a ].}

\subsec{Ghost twisting}

At fixed ghost number, the states in the cohomology of $Q$ resemble the states
of the topological string. For example, the states at ghost-number one are
$V = (\t \g^m \l) A_m(x)$ where $\p_m A_n - \p_n A_m=0$ and $A_m \neq \p_m \L$ for
any $\L$. On a surface of trivial topology, this ghost-number one cohomology vanishes.
However, note that $Q$ is invariant under the scale transformation generated by
\eqn\scale{\Phi = \l^\a w_\a - \t^\a p_\a - u^\a v_\a}
which transforms
\eqn\trans{\l^\a \to \L \l^\a, \quad \t^\a \to \L^{-1}\t^\a, \quad u^\a \to \L^{-1} u^\a,}
$$
w^\a \to \L^{-1} w^\a, \quad p_\a \to \L p_\a, \quad v_\a \to \L v_\a.$$
So one can define a twisted ghost number 
\eqn\twistg{\tilde G = G+\Phi = \l^\a w_\a + u^\a v_\a}
where $G= \t^\a p_\a + 2 u^\a v_\a$ is the original ghost number.
With respect to $\tilde G$, $Q$ still carries $+1$ ghost number but $(\t^a,p_\a)$
now carry zero ghost number.

At fixed twisted ghost number, the cohomology of $Q$ contains non-topological states
describing the super-Maxwell spectrum. These states are described by the vertex operator
\eqn\superm{V = \l^\a A_\a (x,\t) ~\d(u-\l)}
where 
\eqn\deltah{\d(u-\l) \equiv (\l^3 h^{11}) 
\Pi_{I=1}^{11} \d (h_{I\b} (u^\b - \l^\b)), }
$$ (\l^3 h^{11}) \equiv 
\e^{\a_1 ...\a_{16}} h_{1 \a_1} ... h_{11 \a_{11}} (\l\g^m)_{\a_{12}}
(\l\g^n)_{\a_{13}}
(\l\g^p)_{\a_{14}} (\g_{mnp})_{\a_{15}\a_{16}} $$
and $h_{I \a}$ are any 11 spinors such that $(\l^3 h^{11})$ is nonzero.
Note that $\d(u-\l)$ of \deltah\ is independent of the choice of $h_{I\a}$
since it is invariant under
\eqn\gaugeh{\d h_{I\a} = (\g_m\l)_\a \Lambda^m_I}
for any $\L^m_I$.

To verify that $V$ describes the super-Maxwell spectrum, define $U^\a \equiv u^\a
- \l^\a$ where $U^\a$ satisfies $U\g^m\l=0$. Then using
$P_m = -i \p_m$ and $p_\a = {\p\over{\p\t^\a}}$,
$Q= \l^\a D_\a + U^\a {\p\over{\p\t^\a}}$
and $V= \l^\a A_\a (x,\t) \d(U)$ where $D_\a = 
{\p\over{\p\t^\a}} -i (\g^m\t)_\a \p_m$. $QV=0$ implies the super-Maxwell equations
$\g_{m_1 ... m_5}^{\a\b} D_\a A_\b =0$ and $\d V = Q [ \O(x,\t) \d(U)]$ implies
the super-Maxell gauge transformation $\d A_\a = D_\a \O$ where $A_\a(x,\t)$ is
the super-Maxwell spinor gauge superfield \ref\howe{P.S. Howe, {\it Pure spinors
lines in superspace and ten-dimensional supersymmetric theories}, Phys. Lett.
B258 (1991) 141.}. 
So even though the classical action
contains no fermionic variables, the BRST cohomology after twisting the ghost-number is
spacetime supersymmetric and desribes super-Maxwell.

Note that
there are other states in the BRST cohomology in addition to \superm. For example,
the states
\eqn\other{
( \l{\p\over{\p\l}} - u{\p\over{\p u}}- \t{\p\over{\p\t}}) V {\rm ~~and ~~}
[2\l\g^{mn}{\p\over{\p u}} - \l\g^{mn}{\p\over{\p\l}} - 
u\g^{mn}{\p\over{\p u}}+ \t\g^{mn}{\p\over{\p\t}} + i(\t\g^{mnp}\t)\p_p] V}
are also in the cohomology where $V$ is defined in \superm.
However, as will be discussed in the last section, these states are eliminated
after truncating to a ``small'' Hilbert space.

\newsec{Superstring}

\subsec{Worldsheet action and BRST operator}

In this section, the gauge-fixing procedure will be repeated for
the superstring. The classical worldsheet action is constructed
from the variables $x^m$ and the left and right-moving
pure spinors
$\l^\a$ and $\lh^\ah$ satisfying $\l\g^m\l = \lh\g^m\lh=0$
where $\l^\a$ and $\lh^\ah$ have the same spacetime chirality for the Type IIB
superstring and opposite spacetime chirality for the Type IIA superstring.
In addition to the left and right-moving twistor-like constraints
$\p x_m (\g^m\l)_\a =0$ and
$\bar\p x_m (\g^m\lh)_\a =0$ which replace the left and right-moving
Virasoro constraints, one will also need to include the left and
right-moving constraints
$ \p\l^\a =0$ and $\bar\p\lh^\ah=0$. These new constraints are
necessary to close the first-class algebra since 
\eqn\algebra{[ \p x_m (\g^m\l)_\a, 
\p x_n (\g^n\l)_\b] = (\g_{m}\l)_{[\a} (\g^{m}\p\l)_{\b]}, }
$$[ \bar\p x_m (\g^m\lh)_\ah, 
\bar\p x_n (\g^n\lh)_\bh] = (\g_{m}\lh)_{[\ah} (\g^{m} \bar\p\lh)_{\bh]}.$$

The classical worldsheet action is defined as
\eqn\stringact{S_{classical}
 = \int d^2 z [\p x^m \bar\p x_m + w_\a \bar\p \l^\a
+ \hat w_\ah \p \lh^\ah}
$$ + f^\a \p x_m (\g^m \l)_\a + g_\a \p\l^\a + \hat f^\ah
\bar\p x_m (\g^m \lh)_\ah +\hat g_\ah \bar\p\lh^\ah ]$$ 
where $\p= {\partial\over{\partial z}}$ and
$\bar\p= {\partial\over{\partial \bar z}}$,
$f^\a$ and $\hat f^\ah$ are Lagrange multipliers for the twistor-like
constraints $\p x_m (\g^m\l)_\a=0$ and
$\bar\p x_m (\g^m\lh)_\ah=0$, and $g_\a$ and $\hat g_\ah$ are
Lagrange multipliers for the constraints $\p\l^\a=0$ and 
$\bar\p\lh^\ah=0$.

The first step in quantizing this action is to gauge-fix
the Lagrange multipliers
$f^\a = \hat f^\ah = g_\a = \hat g_\ah=0$. This introduces
the fermionic Faddeev-Popov ghosts $(\t^\a, p_\a)$, $(\th^\ah, \hat p_\ah)$,
$(c_\a, b^\a)$ and $(\hat c_\ah, \hat b^\ah)$ with the worldsheet
action
\eqn\faddeev{S_{ghost}
= \int d^2 z ( p_\a \bar\p \t^\a + b^\a \bar\p c_\a
+ \hat p_\ah \bar\p \th^\ah + \hat b^\ah \p \hat c_\ah).}
The resulting left-moving BRST operator is
\eqn\leftbrst{ Q = \int dz [\t^\a \p x_m (\g^m\l)_\a + c_\a \p\l^\a
+ \half (b\g^m\t)(\l\g_m\t) ]}
where the last term comes from the constraint algebra of \algebra.
The right-moving BRST operator is similarly constructed from the
right-moving variables which will sometimes be suppressed. 

As in the superparticle, the constraints $\p x_m (\g^m \l)_\a =0$ are
not all independent and require the introduction of left-moving ghost-for-ghosts
$(u^\a,v_\a)$ satisfying the constraints
\eqn\consuu{u\g^m\l =0.}
Furthermore, only 11 of the 16 components of the $\p\l^\a=0$ constraint 
are independent since $\l\g^m\p\l=0$. To remove the unnecessary ghosts, $(b^\a,c_\a)$
will be required to satisfy a similar constraint
\eqn\consbu{b\g^m\l =0,}
so that only 11 of the 16 $b^\a$'s are independent.
The worldsheet action and left-moving BRST operator
are modified by these ghost-for-ghosts to
\eqn\actiongh{S = 
  \int d^2 z [\p x^m \bar\p x_m + w_\a \bar\p \l^\a
+ \hat w_\ah \p \lh^\ah}
$$+ 
p_\a \bar\p \t^\a + b^\a \bar\p c_\a + v_\a \bar\p u^\a
+ \hat p_\ah \bar\p \th^\ah + \hat b^\ah \p \hat c_\ah + \hat v_\ah \p \hat u^\ah],$$
\eqn\brsttwo{Q = \int dz [\t^\a \p x_m (\g^m\l)_\a + c_\a \p\l^\a
+ \half (b\g^m\t)(\l\g_m\t) + u^\a p_\a ].}

\subsec{Ghost twisting}

As in the superparticle, the BRST operator is invariant under global scale
transformations generated by
\eqn\scalep{\Phi = \int dz (\l^\a w_\a - \t^\a p_\a - u^\a v_\a - c_\a b^\a)}
which transform
\eqn\transp{\l^\a \to \L \l^\a, \quad \t^\a \to \L^{-1}\t^\a, \quad u^\a \to \L^{-1} u^\a,
\quad c_\a \to \L^{-1} c_\a}
$$
w^\a \to \L^{-1} w^\a, \quad p_\a \to \L p_\a, \quad v_\a \to \L v_\a, \quad
b^\a \to \L b^\a.$$
To obtain a nontrivial cohomology, physical states will be defined to have
fixed twisted ghost-number with respect to $\tilde G$ where
\eqn\twistsg{\tilde G = G + \Phi = \int dz (\l^\a w_\a + u^\a v_\a),}
and $G = \int dz (\t^\a p_\a + c_\a b^\a + 2 u^\a v_\a)$ is the original ghost number.

To relate $Q$ of \brsttwo\ with the pure spinor BRST operator, perform the similarity
transformation $Q\to e^{-\int dz (c \p\t)} Q e^{\int dz (c\p\t)}$ so that 
\eqn\brstthr{Q = \int dz [ (\l\g_m\t)\p x^m + \half (\p\t\g^m\t)(\l\g_m\t)
+u^\a p_\a + c_\a (\p\l^\a-\p u^\a) + \half(b\g^m\t)(\l\g_m\t)]}
$$= 
 \int dz [ \l^\a d_\a +
U^\a (p_\a +\p c_\a)  + \half(b\g^m\t)(\l\g_m\t)]$$
where
\eqn\defUd{U^\a = u^\a -\l^\a, \quad d_\a = p_\a + \p x^m (\g_m\t)_\a - 
\half (\g_m\t)_\a (\t\g^m\p\t).}
After performing the similarity transformation, the constraint of \consbu\
needs to be modified since it no longer anticommutes with $Q$. A suitable
modified constraint is 
\eqn\bconst{ b\g^m\l = U\g^m (b+\p\t), }
which is BRST-invariant with respect to \brstthr\ and which coincides with
the original constraint when $U^\a=0$.

\subsec{Physical states}

In analogy with the super-Maxwell vertex operator $V = \l^\a A_\a(x,\t) \d(U)$
of the previous section, the naive guess for the general vertex operator is
\eqn\naive{V = V_0 (\l,x,\t,p,w) \d(U)}
where $V_0$ is in the cohomology of $\int dz \l^\a d_\a$.
However, this is not quite right since the terms $\int dz~ U^\a p_\a$ and
$\int dz~\half (b\g^m\t)(\l\g_m\t)$ in $Q$ might not annihilate $V$ if $V_0$ 
involves $\p \t^\a$ or $p_\a$.

To ensure that the vertex operator is annihilated by $Q$, consider
\eqn\massiveone{V = V_0 (\l,x,\t,p,w) Y \d(U) }
where $Y \d(U)$ is 
defined by
\eqn\defY{Y \d(U)= \prod_{I=1}^{11} (h_{I\a} b^\a) \d(h_{I\rho} U^\rho)
\d(h_{I\b} \p U^\b).}
As before, $Y \d(U)$ is independent of the choice of $h_{I\a}$ since
it is invariant under \gaugeh. Furthermore, $QV=0$ for all states 
where $p_\a$ has at most a double pole and
$\half (\g^m\t)_\a(\l\g_m\t)$ has at most a single pole with $V_0$.

To describe a general massive state where $p_\a$ and 
$\half (\g^m\t)_\a(\l\g_m\t)$ can have more singular poles,
define
\eqn\massivegen{V = V_0 (\l,x,\t,p,w) \lim_{n\to\infty} Y^n \d(U) }
where 
\eqn\defYn{Y^n \d(U)= \prod_{I=1}^{11} (h_{I\a} b^\a) ... (h_{I\b}\p^{n-1} b^\b)
~ \d(h_{I\rho} U^\rho)\d(h_{I\g} \p U^\g) ...
\d(h_{I\d} \p^n U^\d) .}
In other words, 
\eqn\otherw{V = V_0(\l,x,\t,p,w)|0\rangle}
where $|0\rangle$ is annihilated by all positive and negative modes
of $U^\a$ and $b^\a$ and is not annihilated by any modes of $v_\a$ or
$c_\a$. Note that $|0\rangle$ carries zero conformal weight because
of cancellation between the bosonic variables $(U^\a, v_\a)$ 
and fermionic variables $(b^\a, c_\a)$.

\newsec{Scattering Amplitudes}

\subsec{Non-minimal variables}

As in the pure spinor formalism, scattering amplitude computations are simplified by
introducing ``non-minimal" left-moving bosonic worldsheet variables $\bar\l_\a$ satisfying
$\bar\l\g^m\bar\l=0$ and its conjugate momentum $\bar w^\a$. 
Formally, $\bar\l_\a$ can be interpreted as the complex
conjugate of $\l^\a$ in Euclidean signature, and it is natural to identify both
$\l^\a$ and $\bar\l_\a$ as classical variables. One should also introduce the
classical constraint $\bar w^\a=0$ so that $\bar\l_\a$ has no effect on the
cohomology.

The classical worldsheet action involving the left and right-moving non-minimal
variables is
\eqn\actnonm{
S = 
  \int d^2 z [\p x^m \bar\p x_m + w_\a \bar\p \l^\a
+ \hat w_\ah \p \lh^\ah + \bar w^\a \bar\p \bar\l_\a + \hat{\bar w}^\ah \p \hat{\bar\l}_\ah}
$$ +e_\a \bar w^\a + f^\a \p x_m (\g^m \l)_\a + g_\a \p\l^\a +\hat e_\ah \hat w^\ah +
 \hat f^\ah
\bar\p x_m (\g^m \lh)^\ah +\hat g_\ah \bar\p\lh^\ah ]$$ 
where the Lagrange multipliers $e_\a$ and $\hat e_\ah$ are constrained to satisfy
$e\g^m\bar\l =0$ and $\hat e\g^m\hat{\bar\l}=0$.
Gauge-fixing $e_\a=0$ and $\hat e_\ah=0$ introduces the fermionic Faddeev-Popov ghosts
$(r_\a,s^\a)$ and $(\hat r_\ah, \hat s^\ah)$ satisfying the constraints
$r\g^m\bar\l=0$ and $\hat r \g^m\hat{\bar\l}=0$ with the worldsheet action
\eqn\actfp{\int d^2 z (s^\a \bar\p r_\a + \hat s^\ah \p \hat r_\ah)}
and modifies the left-moving BRST operator to
\eqn\brstnonm{Q = \int dz [\t^\a \p x_m (\g^m\l)_\a + c_\a \p\l^\a
+ \half (b\g^m\t)(\l\g_m\t) + u^\a p_\a + \bar w^\a r_\a ].}

\subsec{Small Hilbert space}

Since $c_\a$ only appears with derivatives in the BRST operator,
it is consistent to remove its zero mode from the Hilbert space as
one does with the $\xi$ zero mode of the RNS bosonized ghosts
\ref\fms{D. Friedan, E. Martince and S. Shenker, {\it Conformal
invariance, supersymmetry and string theory}, Nucl. Phys. B271 (1986) 93.}.
So physical states $V$ will be required to be in the ``small'' Hilbert
space, i.e. they need to satisfy $b_0^\a V=0$.

This requirement is necessary since, as in the pure spinor formalism,
the measure factor for tree level amplitudes will be defined as
$\langle (\l^3 \t^5)\rangle =1$ where 
\eqn\lcubed{(\l^3\t^5) \equiv (\l\g^m\t)(\l\g^n\t)(\l\g^p\t)(\t\g_{mnp}\t).}
For amplitudes to be BRST invariant, $(\l^3\t^5)$ must be in the cohomology
of $Q$. However, $(\l^3\t^5) = Q\Omega$ where $\Omega= 
(c\g^{mn}\l)(\l\g^p\t)(\t\g_{mnp}\t)$. Since $b_0^\a \Omega$ is
nonzero, this does not cause problems if one restricts physical states
to the small Hilbert space.

Interestingly, the ``picture-raising'' operator associated with the $c_\a$
zero mode is 
\eqn\pic{\{Q, c_\a \} = \half (\g^m\t)_\a (\l\g^m\t)}
which is the unintegrated vertex operator for the zero-momentum gluino.
The integrated vertex operator of this state is the spacetime supersymmetry
generator, so ``picture-raising" is related to spacetime supersymmetry. 

In the next subsection, it will be argued that the path integrals
over the bosonic $(U^\a, v_\a)$ and fermionic
$(b^\a,c_\a)$ variables should cancel
each other. In other for these path integrals to cancel, one needs
to impose a second restriction on states in the ``small'' Hilbert
space that they satisfy $U_0^\alpha V=0$ in addition to $b_0^\a V=0$.
This second restriction truncates out the states of \other\ and
reduces the cohomology to the superstring spectrum.

\subsec{Tree amplitude prescription}

The states of \massivegen\ are constructed out  
of a ground state $|0\rangle$ annihilated by both positive and
negative modes of $U^\a$ and $b^\a$. So the path integral over
the $(U^\a,v_\a)$ and $(b^\a,c_\a)$ variables is not the usual one.
However, since the operators $V_0(\l,x,\t,p,w)$ appearing in
vertex operators are
independent of the $(U^\a,v_\a)$ and $(b^\a, c_\a)$ variables,
one never needs to evaluate correlation functions for these variables
and only needs to know their partition functions. It will be assumed
that the partition function of the bosonic $(U^\a,v_\a)$ variables
cancels the partition function of the fermionic $(b^\a,c_\a)$ variables,
so the scattering amplitude computation reduces to the path integral
over the pure spinor non-minimal variables $(x,\t,\l,\bar\l,r, p,w,\bar w, s)$.

As shown in \ref\ntwo{N. Berkovits, {\it Pure spinor formalism as an N=2
topological string}, JHEP 0510 (2005) 089, hep-th/0509120.}, 
the path integral over non-minimal variables reproduces the tree-level 
measure factor of $\langle (\l^3\t^5)\rangle =1$.
So the N-point tree amplitude prescription is simply
\eqn\amp{\langle V_0^{(1)}(z_1)
V_0^{(2)}(z_2) V_0^{(3)}(z_3) \prod_{r=4}^N \int dz_r U_0^{(r)}(z_r)\rangle}
where the integrated vertex operator is
$U^{(r)} = U_0^{(r)} |0\rangle$ and $U_0^{(r)}$ satisfies
\eqn\sating{[ \int \l^\a d_\a, U_0^{(r)}(z_r) ] = \p V_0^{(r)}(z_r).}
Assuming the partition functions over the $(U^\a,v_\a)$ and
$(b^\a,c_\a)$ variables cancel each other, the prescription of \amp\ 
therefore
reproduces the usual pure spinor tree amplitude prescription.

\vskip 15pt

{\bf Acknowledgements:} I would like to thank Ido Adam,
Andrei Mikhailov, Nikita Nekrasov, Cumrun Vafa and Edward Witten for
useful discussions, 
CNPq grant 300256/94-9 and FAPESP grant 09/50639-2
for partial financial support, and the KITP program ``The Harmony
of Scattering Amplitudes'' 
for their hospitality.

\listrefs

\end